\documentclass[aps,pre,onecolumn,superscriptaddress]{revtex4}
\usepackage{amsmath}
\usepackage{graphicx}
\usepackage{fancybox}
\usepackage{subfigure}

\graphicspath{{./Figures/}}

\newcommand{\eref}[1]{Eq.~(\ref{#1})}

\newcommand{\vect}[1]{\mathbf{#1}}

\newcommand{\vdiff}[2]{\left|\vect{#1} - \vect{#2}\right|}

\newcommand{\eg}{\emph{e.g.}}

\newcommand{\allpos}{\vect{\overline{R}}}

\newcommand{\kT}{\ensuremath{{\rm k_BT}}}

\begin{document}


\title{Exact relations between
charge-density functions determining the total Coulomb energy and the dielectric constant for a mixture of 
neutral and charged site-site molecules}


\author{Jocelyn M. Rodgers}
\altaffiliation[Present address: ]{Materials Science Division, Lawrence Berkeley National Laboratory, Berkeley, CA 94720}
\affiliation{Institute for Physical Science and Technology, University of
  Maryland, College Park, Maryland 20742}
\affiliation{Chemical Physics Program, University of Maryland, College Park, Maryland 20742}

\author{John D. Weeks}
\email{jdw@ipst.umd.edu}
\affiliation{Institute for Physical Science and Technology, University of
  Maryland, College Park, Maryland 20742}
\affiliation{Department of Chemistry and Biochemistry, University of Maryland,
  College Park, Maryland 20742}


\date{\today}

\pacs{}


\begin{abstract}
  We extend results developed
  by Chandler~[J.~Chem.~Phys.\ {\bf 65}, 2925 (1976)] for the dielectric constant
  of neutral site-site molecular models to mixtures of both charged and uncharged
  molecules. This provides a unified derivation connecting the Stillinger-Lovett moment conditions
  for ions to standard results for the dielectric constant for polar species
  and yields exact expressions
  for the small-$k$ expansion of the two-point intermolecular charge-density function used
to determine the total Coulomb energy. The latter is useful in determining corrections to the thermodynamics
 of uniform site-site molecular models simulated with spherically truncated Coulomb interactions.
 \end{abstract}
 
 \maketitle
  
In this note we extend results for the small wavevector expansion
small-$k$ relationships between the dielectric constant and moments of
the intermolecular and intramolecular correlations functions,
originally developed by
Chandler~\cite{Chandler.1977.Dielectric-constant-and-related-equilibrium-properties-of-molecular}.
This extension generalizes these relations to mixtures of
neutral and charged molecular species and connects them to the
small wavevector expansion of the two-point intermolecular charge-density function used
to determine the total Coulomb energy. These results are used elsewhere to derive corrections
to the thermodynamics of uniform site-site molecular models simulated
with spherically truncated Coulomb interactions~\cite{dipolethermo}.

\section{Derivation of Zeroth and Second Moment Conditions for a Mixture of Neutral and Charged Site-Site Molecules}

The total Coulomb energy obtained during simulation of a mixture of
small site-site molecular species without any intramolecular charge-charge interactions is
\begin{equation}
  U^q = \left< \frac{1}{2} \sum_{M} \sum_{M^\prime} \sum_{i=1}^{N_M}
    \sum_{j=1}^{N_{M^\prime}} \left(1 - \delta_{MM^\prime}\delta_{ij}
    \right) \sum_{\alpha=1}^{n_M} \sum_{\gamma=1}^{n_{M^\prime}}
    \frac{q_{\alpha M} q_{\gamma M^\prime}}{\left| \vect{r}_{i
          M}^{(\alpha)} - \vect{r^\prime}_{jM^\prime}^{(\gamma)} \right|}\right>.
  \label{eqn:energysum}
\end{equation}
In this notation, the angular brackets indicate a normalized ensemble
average, $M$ and $M^\prime$ indicate a given molecular species, $i$
and $j$ indicate a given molecule of a given species, and $\alpha$ and
$\gamma$ represent the intramolecular
sites~\cite{ChandlerPratt.1976.Statistical-Mechanics-of-Chemical-Equilibria-and-Intramolecular,Chandler.1977.Dielectric-constant-and-related-equilibrium-properties-of-molecular}.
The Kronecker deltas are necessary to exclude any charge-charge interactions between intramolecular sites
within a given molecule.  This energy $U^q$ can be more compactly represented as
\begin{equation}
  U^q = \frac{1}{2} \int d\vect{r} \int d\vect{r^\prime}
  \frac{\rho^{qq}(\vect{r}, \vect{r^\prime})}{\vdiff{r}{r^\prime}},
  \label{eqn:rhoqq}
\end{equation}
where $\rho^{qq}$ is a two-point intermolecular charge-density function that
explicity excludes any purely intramolecular charge correlations, as
implied by \eref{eqn:energysum} and detailed below.

The composite function $\rho^{qq}(\vect{r},\vect{r}^\prime)$ is a
charge-weighted linear combination of all intermolecular, two-point
site-site distribution functions,
\begin{equation}
  \rho^{qq}(\vect{r},\vect{r}^\prime) \equiv \sum_{\alpha M} \sum_{\gamma M^\prime} q_{\alpha M}
  q_{\gamma M^\prime} \rho_{\alpha M \gamma M^\prime} (\vect{r},\vect{r}^\prime).
\end{equation}
We now relate this function to the basic charge-charge linear response function used in the theory of the dielectric constant.

For solutions of primitive model ions, charge neutrality and screening
place specific requirements on the behavior of $\rho^{qq}$ in
$k$-space at small $k$
values~\cite{StillingerLovett.1968.General-Restriction-on-Distribution-of-Ions-in-Electrolytes}.
More generally, for a fluid composed of charged and polar molecules,
the dielectric screening behavior of the molecules places restrictions
on the decay of this two-point charge density. Based on this
observation, we are able to harness a theoretical development of
Chandler~\cite{Chandler.1977.Dielectric-constant-and-related-equilibrium-properties-of-molecular}
that expresses the dielectric constant in terms of an exact sum of
charge-density-weighted pair correlation functions. We generalize the
derivation to include both charged and neutral site-site molecules and
we take the dielectric constant as a given. From this vantage point,
we may instead use these relations to place requirements on the decay
of the two-point charge density $\rho^{qq}$.

We first define the instantanteous single-point charge-density
$\rho^q(\vect{r},\allpos)$, a function of both a given
external position $\vect{r}$ and the set of positions of all mobile particles
$\allpos \equiv \left\{ \vect{R}_{iM} \right\} \equiv \left\{ \vect{r}_{iM}^{(\alpha)} \right\} $, as
\begin{equation}
  \rho^q (\vect{r},\allpos) \equiv \sum_{M} \sum_{i=1}^{N_M}
  \sum_{\alpha=1}^{n_M} q_{\alpha M} \, \delta ( \vect{r} - \vect{r}_{iM}^{(\alpha)} ).
  \label{eqn:rhoqdef}
\end{equation}
With such a definition, the ensemble-averaged charge-density profile $\rho^q(\vect{r})$ is
\begin{equation}
  \rho^q(\vect{r}) = \left< \rho^q (\vect{r},\allpos) \right>.
\end{equation}
In the case of a uniform system, $\rho^q(\vect{r}) = 0$.
Comparing Eqs.\ (\ref{eqn:energysum}) and (\ref{eqn:rhoqq}) and using Eq.\ (\ref{eqn:rhoqdef}),
we may also express $\rho^{qq}(\vdiff{r}{r^\prime})$ for a
uniform system as
\begin{equation}
\rho^{qq}(\vdiff{r}{r^\prime}) = \left< \rho^q (\vect{r},\allpos)
  \rho^q (\vect{r^\prime},\allpos)\right> - \left< \sum_{M} N_M \sum_{\alpha=1}^{n_M}
  \sum_{\gamma=1}^{n_M} q_{\alpha M} q_{\gamma M} \delta ( \vect{r}-\vect{r}_{1M}^{(\alpha)}) \delta ( \vect{r^\prime}-\vect{r}_{1M}^{(\gamma)} )\right>.
\end{equation}
We have used the equivalence of all molecules of type $M$ in the last term.
This term removes purely intramolecular charge-density correlations;
we shall determine the small-$k$ contributions from that term based on
well-known molecular properties using the approach of
Chandler~\cite{Chandler.1977.Dielectric-constant-and-related-equilibrium-properties-of-molecular}
later in this note.  The first term, in contrast, is exactly the
charge-charge linear response function for a uniform neutral system:
\begin{equation}
  \left< \rho^q(\vect{r},\allpos) \rho^q
    (\vect{r^\prime},\allpos) \right> = \left<
    \delta\rho^q(\vect{r},\allpos) \delta \rho^q
    (\vect{r^\prime},\allpos) \right> \equiv \chi^{qq}(\vdiff{r}{r^\prime}).
  \label{eqn:chiqquniform}
\end{equation}
Here $\delta \rho^q(\vect{r},\allpos) \equiv \rho^q(\vect{r},\allpos) - \left< \rho^q (\vect{r},\allpos) \right>$.
Physically $\chi^{qq}$ describes the coupling between charge-density
fluctuations at positions $\vect{r}$ and $\vect{r^\prime}$. As is well
established~\cite{Chandler.1977.Dielectric-constant-and-related-equilibrium-properties-of-molecular,HansenMcDonald.2006.Theory-of-Simple-Liquids},
such a function is intimately related to the dielectric behavior of
the fluid at long distances, and furthermore, may be easily analyzed
using basic electrostatics and standard definitions of the
functional derivative.

The electrostatic potential at $\vect{r}$ induced by a fixed external
charge distribution $\rho^q_{\rm ext}(\vect{r^\prime})$ (\eg, a test
charge $Q$ placed at the origin, as considered by Chandler) is given
by
\begin{equation}
  \mathcal{V}_{\rm ext}(\vect{r}) = \int \frac{\rho^q_{\rm ext}(\vect{r^\prime})}{\vdiff{r}{r^\prime}} d \vect{r}^\prime,
\end{equation}
and the associated electrostatic energy for a particular microscopic configuration characterized by the set of molecular positions $\allpos$ is then
\begin{equation}
  U^q_{\rm ext}(\allpos) = \int \rho^q(\vect{r},\allpos) \mathcal{V}_{\rm ext}(\vect{r}) d \vect{r}.
\end{equation}
This energy contribution will appear in the nonuniform system's Hamiltonian when $\mathcal{V}_{\rm ext}(\vect{r})$ is nonzero.
As such, we know from standard definitions of functional
differentiation of free
energies~\cite{HansenMcDonald.2006.Theory-of-Simple-Liquids,Percus.J.The-pair-distribution-function-in-classical-statistical.1964}
that
\begin{equation}
  \frac{\delta \left[ -\beta A \right]}{\delta \left[ -\beta
      \mathcal{V}_{\rm ext}(\vect{r})\right]} = \left< \rho^q ( \vect{r},
    \allpos)\right>_{\mathcal{V}} \equiv  \rho^q_{\mathcal{V}}(\vect{r}) ,
\end{equation}
where $\beta\equiv(\kT)^{-1}$ and the subscript $\mathcal{V}$ indicates that the ensemble average is taken in the presence of an external potential.
Similarly we have
\begin{equation}
  \frac{\delta \rho^q_{\mathcal{V}}(\vect{r})}{\delta \left[ -\beta
      \mathcal{V}_{\rm ext}(\vect{r}^\prime) \right]} =   \frac{\delta \left[ -\beta A \right] }{\delta \left[ -\beta
      \mathcal{V}_{\rm ext}(\vect{r}) \right] \delta \left[ -\beta
      \mathcal{V}_{\rm ext}(\vect{r}^\prime) \right]} = \chi^{qq}_{\mathcal{V}}\left(
    \vect{r}, \vect{r^\prime}\right).
  \label{eqn:chiv}
\end{equation}

The total electrostatic potential at position $\vect{r}$ in the nonuniform fluid is then given by the sum of the external potential and the induced polarization potential:
\begin{equation}
  \mathcal{V}_{\rm tot}(\vect{r}) = \mathcal{V}_{\rm ext}(\vect{r}) +
  \mathcal{V}_{\rm pol}(\vect{r}) = \mathcal{V}_{\rm ext}(\vect{r}) + \int d\vect{r}^\prime \,
  \frac{\rho^{q}_{\mathcal{V}}(\vect{r}^\prime)}{\vdiff{r}{r^\prime}}.
\end{equation}
To get a formula for the dielectric constant we expand about the
uniform neutral system and evaluate $\mathcal{V}_{\rm pol}$ to linear order in
$\mathcal{V}_{\rm ext}$ using \eref{eqn:chiv} and find
\begin{align}
  \mathcal{V}_{\rm tot}(\vect{r}) &\approx \mathcal{V}_{\rm
    ext}(\vect{r}) + \int d\vect{r}^\prime \,
  \frac{1}{\vdiff{r}{r^\prime}}  \int d\vect{r}^{\prime \prime} \, \frac{\delta \rho^{q}(\vect{r}^\prime)}{\delta \left[ -\beta
     \mathcal{V}_{\rm ext}(\vect{r}^{\prime \prime}) \right]} \left[ -\beta \mathcal{V}_{\rm
  ext} (\vect{r}^{\prime \prime})\right], \\
  &= \mathcal{V}_{\rm ext}(\vect{r}) - \int d\vect{r}^\prime \,
  \frac{1}{\vdiff{r}{r^\prime}}  \int  d\vect{r}^{\prime \prime} \, \beta \chi^{qq} \left(
    \vdiff{r^\prime}{r^{\prime \prime}} \right) \mathcal{V}_{\rm
  ext} (\vect{r}^{\prime \prime}).
\end{align}
Here $\chi^{qq}$ is the linear response function in the uniform fluid as in \eref{eqn:chiqquniform}.
Taking the Fourier transform of the final equation, we find
\begin{equation}
  \hat{\mathcal{V}}_{\rm tot}(\vect{k}) = \hat{\mathcal{V}}_{\rm ext}(\vect{k})- \frac{4
    \pi}{k^2} \beta \hat{\chi}^{qq}(k) \hat{\mathcal{V}}_{\rm ext}(\vect{k}).
\end{equation}
Thus to linear order we have
\begin{equation}
 \frac{\hat{\mathcal{V}}_{\rm tot}(\vect{k})}{\hat{\mathcal{V}}_{\rm ext}(\vect{k})} = 1 -
 \frac{4 \pi \beta }{k^2} \hat{\chi}^{qq}(k).
\end{equation}

Phenomenologically, we know that in the limit of $\vect{k} \rightarrow
0$, this ratio of the total electrostatic potential to the
externally-imposed potential is exactly $1/\epsilon$. Therefore, we
find for our molecular mixture the general result
\begin{equation}
  \lim_{\vect{k} \rightarrow 0} \left( 1 - \frac{4 \pi \beta}{k^2}
    \hat{\chi}^{qq}(k) \right) = \frac{1}{\epsilon}.
  \label{eqn:ChiLimit}
\end{equation}
Based on the limit in \eref{eqn:ChiLimit}, and expanding $\hat{\chi}^{qq}$ for small $k$
as $\hat{\chi}^{(0)qq} + \hat{\chi}^{(2)qq}k^2$, we have 
\begin{align}
  \hat{\chi}^{(0)qq} &= 0 \nonumber \\
  4 \pi \beta \hat{\chi}^{(2)qq} &= 1 - \frac{1}{\epsilon}.
  \label{eqn:chi0}
\end{align}
Any mixture with mobile ions acts as a conductor with $\epsilon = \infty$ in Eq.\ (\ref{eqn:chi0}),
independent of the nature of the neutral components.

Our goal is to write a small-$k$ expansion of the two-point intermolecular  charge density,
\begin{equation}
\hat{\rho}^{qq}(k) \approx \hat{\rho}^{(0)qq} + k^2 \hat{\rho}^{(2)qq} + \mathcal{O}(k^4).
\end{equation}
As stated at the beginning of this derivation, 
\begin{equation*}
\rho^{qq}(\vdiff{r}{r^\prime}) = \chi^{qq}\left( \vdiff{r}{r^\prime} \right)- \left< \sum_{M} N_M \sum_{\alpha=1}^{n_M}
  \sum_{\gamma=1}^{n_M} q_{\alpha M} q_{\gamma M} \delta ( \vect{r}-\vect{r}_{1M}^{(\alpha)}
  ) \delta ( \vect{r^\prime}-\vect{r}_{1M}^{(\gamma)} )\right>.
\end{equation*}
We already have results for the small-$k$ expansion of the
charge-charge linear response function $\hat{\chi}^{qq}$. Now we must
remove the intramolecular contributions as described by the second
term in the equation above. Defining the conditional singlet
intramolecular site density functions $\varrho_{\alpha | \gamma
  M}(\vect{r}| \vect{r^\prime})$ for $\alpha \neq \gamma$ as
\begin{equation}
  \rho_{\gamma M}(\vect{r^\prime}) \varrho_{\alpha | \gamma M}(\vect{r}|\vect{r^\prime}) = \left< {N_M} \delta ( \vect{r}-\vect{r}_{1M}^{(\alpha)}
  ) \delta ( \vect{r^\prime}-\vect{r}_{1M}^{(\gamma)} ) \right>,
\end{equation}
 and applying consequences of uniformity, we find
\begin{equation}
  \rho^{qq}(\vdiff{r}{r^\prime}) = \chi^{qq}\left( \vdiff{r}{r^\prime}
  \right)- 
  \sum_{M} \rho_M \sum_{\alpha, \gamma} q_{\alpha M} q_{\gamma M} \left[ \delta_{\alpha \gamma}
  \delta(\vect{r}-\vect{r}^\prime) + \varrho_{\alpha | \gamma M}(\vdiff{r}{r^\prime}) \right].
\end{equation}
Using the notation of Chandler for the term in square
brackets~\cite{Chandler.1977.Dielectric-constant-and-related-equilibrium-properties-of-molecular}, we may write this expression more compactly as
\begin{equation}
  \rho^{qq}(\vdiff{r}{r^\prime}) = \chi^{qq}\left( \vdiff{r}{r^\prime}
  \right)- 
  \sum_{M} \rho_M \sum_{\alpha, \gamma} q_{\alpha M} q_{\gamma M} \omega_{\alpha \gamma M}(\vdiff{r}{r^\prime}).
  \label{eqn:rhoqqunif}
\end{equation}
For neutral molecules, Chandler demonstrated that the small-$k$
components of $\hat{\omega}_{\alpha \gamma M}(k)$ are related to simple
properties of the molecule.   For both charged and uncharged molecules, the zeroth
moment of $\hat{\omega}_{\alpha \gamma M}$ is simply
\begin{equation}
  \hat{\omega}_{\alpha \gamma M} ^{(0)} = \delta_{\alpha \gamma} + \int d\vect{r} \,\varrho_{\alpha | \gamma M} (\vect{r}) = \delta_{\alpha \gamma} + (1 - \delta_{\alpha \gamma}) = 1.
\end{equation}
Using this exact expression in Eq.\ (\ref{eqn:rhoqqunif}) yields
\begin{equation}
\hat{\rho}^{(0)qq} = \hat{\chi}^{(0)qq} -  \sum_{M} \rho_M
\sum_{\alpha, \gamma} q_{\alpha M} q_{\gamma M} = - \sum_M \rho_M q_M^2,
\end{equation}
an expression encompassing the standard zeroth moment condition for
ions~\cite{StillingerLovett.1968.General-Restriction-on-Distribution-of-Ions-in-Electrolytes} and the
zeroth moment for neutral molecular
species~\cite{Chandler.1977.Dielectric-constant-and-related-equilibrium-properties-of-molecular}.

The expression for $\hat{\omega}^{(2)}_M$ determined by
Chandler~\cite{Chandler.1977.Dielectric-constant-and-related-equilibrium-properties-of-molecular} may be written most generally as
\begin{equation}
  \hat{\omega}^{(2)}_M \equiv \sum_{\alpha \neq \gamma} q_{\alpha M} q_{\gamma M} \hat{\omega}_{\alpha \gamma M}^{(2)} = - \frac{1}{6} \int d\vect{r} \sum_{\alpha \neq \gamma} q_{\alpha M}  q_{\gamma M} \varrho_{\alpha | \gamma M}(\vect{r}) r^2 = -\frac{1}{6} \sum_{\alpha \neq \gamma} q_{\alpha M} q_{\gamma M} \left< l_{\alpha \gamma M}^2 \right>,
\end{equation}
where $l_{\alpha \gamma M}$ is the bondlength between sites $\alpha$
and $\gamma$ for a molecule of species $M$. As shown in
Ref.~\onlinecite{Chandler.1977.Dielectric-constant-and-related-equilibrium-properties-of-molecular},
for a \emph{neutral} molecule indicated by $N$ below, the final
summation in the above equation is simply related to the molecular dipole moment
$\mu_N$ and the molecular polarizability $\alpha_N$ as
\begin{equation}
\hat{\omega}^{(2)}_N = \sum_{\alpha \neq \gamma} q_{\alpha N} q_{\gamma N} \hat{\omega}_{\alpha \gamma N}^{(2)} =  \frac{1}{3} \mu_N^2 + \kT \alpha_N.
\end{equation}
This relationship does not hold for a charged molecule since the
dipole moment then depends on the choice of coordinate system.

Distinguishing charged species ($C$) and  neutral species ($N$) where $\{ M \} = \{ N \} \cup \{ C \}$, and
without substituting for $\hat{\omega}_N^{(2)}$ and $\hat{\omega}_C^{(2)}$, we find
\begin{equation}
\hat{\rho}^{(2)qq} = \frac{\kT}{4\pi} \frac{\epsilon -
  1}{\epsilon}- \sum_M \rho_M \sum_{\alpha \neq \gamma} q_{\alpha M}
q_{\gamma M} \hat{\omega}_{\alpha \gamma M}^{(2)} = \frac{\kT}{4\pi} \frac{\epsilon -
  1}{\epsilon} - \sum_N \rho_N \hat{\omega}_N^{(2)} - \sum_C \rho_C \hat{\omega}_C^{(2)}.
\end{equation}
Thus, we may write a general expression for $\hat{\rho}^{qq}$ in $k$-space.
Utilizing the expressions for $\hat{\omega}_N^{(2)}$ and $\hat{\omega}_C^{(2)}$, we have
\begin{equation}
\hat{\rho}^{qq}(k) = - \sum_C \rho_C q_C^2 + k^2 \frac{\kT}{4\pi} \frac{\epsilon - 1}{\epsilon} - k^2 \sum_N \rho_N \left\{ \frac{1}{3} \mu_N^2 + \kT \alpha_N \right\} + k^2 \frac{1}{6} \sum_{C} \rho_C \sum_{\alpha \neq \gamma} q_{\alpha C}q_{\gamma C}\left< l_{\alpha \gamma C}^2 \right> + \mathcal{O}\left( k^4 \right).
\end{equation}

Unlike $\chi^{qq}$ in Eq.\ (\ref{eqn:chi0}), we see that the small-$k$
behavior of this two-point intermolecular charge-density function
depends on several simple properties of the solution as a whole, like
the dipole moment and polarizability of individual neutral molecules,
and the net molecular charge and the average square bond lengths of
charged molecules, as well as the dielectric constant. Thus by knowing
simple single molecule properties and the long wavelength dielectric
constant, we know how intermolecular charge-charge correlations decay
in solution. This is the essential equation used to develop energy and
pressure corrections for simulations of bulk liquids using molecular
models with truncated Coulomb interactions~\cite{dipolethermo}.  A
related expression may be developed for larger molecular species with
intramolecular charge-charge interactions.

This work was supported by NSF grant CHE-0517818. JMR acknowledges the
support of the University of Maryland Chemical Physics fellowship.


\begin{thebibliography}{9}
\expandafter\ifx\csname natexlab\endcsname\relax\def\natexlab#1{#1}\fi
\expandafter\ifx\csname bibnamefont\endcsname\relax
  \def\bibnamefont#1{#1}\fi
\expandafter\ifx\csname bibfnamefont\endcsname\relax
  \def\bibfnamefont#1{#1}\fi
\expandafter\ifx\csname citenamefont\endcsname\relax
  \def\citenamefont#1{#1}\fi
\expandafter\ifx\csname url\endcsname\relax
  \def\url#1{\texttt{#1}}\fi
\expandafter\ifx\csname urlprefix\endcsname\relax\def\urlprefix{URL }\fi
\providecommand{\bibinfo}[2]{#2}
\providecommand{\eprint}[2][]{\url{#2}}

\bibitem[{\citenamefont{Chandler}(1977)}]{Chandler.1977.Dielectric-constant-and-related-equilibrium-properties-of-molecular}
\bibinfo{author}{\bibfnamefont{D.}~\bibnamefont{Chandler}},
  \bibinfo{journal}{J. Chem. Phys.} \textbf{\bibinfo{volume}{67}},
  \bibinfo{pages}{1113} (\bibinfo{year}{1977}).

\bibitem{dipolethermo} J.M. Rodgers and J. D. Weeks, (to be published)

\bibitem[{\citenamefont{Chandler and
  Pratt}(1976)}]{ChandlerPratt.1976.Statistical-Mechanics-of-Chemical-Equilibria-and-Intramolecular}
\bibinfo{author}{\bibfnamefont{D.}~\bibnamefont{Chandler}} \bibnamefont{and}
  \bibinfo{author}{\bibfnamefont{L.~R.} \bibnamefont{Pratt}},
  \bibinfo{journal}{J. Chem. Phys.} \textbf{\bibinfo{volume}{65}},
  \bibinfo{pages}{2925} (\bibinfo{year}{1976}).
  
  \bibitem[{\citenamefont{Stillinger and
  Lovett}(1968)}]{StillingerLovett.1968.General-Restriction-on-Distribution-of-Ions-in-Electrolytes}
\bibinfo{author}{\bibfnamefont{F.~H.} \bibnamefont{Stillinger}}
  \bibnamefont{and} \bibinfo{author}{\bibfnamefont{R.}~\bibnamefont{Lovett}},
  \bibinfo{journal}{J. Chem. Phys.} \textbf{\bibinfo{volume}{49}},
  \bibinfo{pages}{1991} (\bibinfo{year}{1968}).


\bibitem[{\citenamefont{Hansen and
  McDonald}(2006)}]{HansenMcDonald.2006.Theory-of-Simple-Liquids}
\bibinfo{author}{\bibfnamefont{J.-P.} \bibnamefont{Hansen}} \bibnamefont{and}
  \bibinfo{author}{\bibfnamefont{I.~R.} \bibnamefont{McDonald}},
  \emph{\bibinfo{title}{Theory of Simple Liquids}}
  (\bibinfo{publisher}{Academic Press}, \bibinfo{address}{New York},
  \bibinfo{year}{2006}), \bibinfo{edition}{3rd} ed.

\bibitem[{\citenamefont{Percus}(1964)}]{Percus.J.The-pair-distribution-function-in-classical-statistical.1964}
\bibinfo{author}{\bibfnamefont{J.~K.} \bibnamefont{Percus}}, 
\bibinfo{article}{The pair distribution function in classical statistical
  mechanics},
  In \emph{\bibinfo{title}{The Equilibrium Theory of Classical Fluids}} (\bibinfo{publisher}{W. A. Benjamin, Inc.}, \bibinfo{address}{New
  York}, \bibinfo{year}{1964}). Our definition of $\chi^{qq}$ is consistent with notation in this reference and
Ref. \cite{Chandler.1977.Dielectric-constant-and-related-equilibrium-properties-of-molecular}.

\end{thebibliography}

\end{document}